\documentclass[preprint,12pt]{elsarticle}

\usepackage{amssymb}

\journal{Applied Surface Science}

\begin{document}

\begin{frontmatter}



\title{An atomically flat single--crystalline gold film thermometer on mica to study energy (heat) exchange at the nano--scale}


\author[label1]{S. Veronesi\corref{cor1}}
\ead{stefano.veronesi@nano.cnr.it}
\author[label1]{T. Papa}
\author[label1]{Y. Murata}
\author[label1]{S. Heun}

\cortext[cor1]{Corresponding author}

\address[label1]{NEST, Istituto Nanoscienze--CNR and Scuola Normale Superiore, Piazza S. Silvestro 12, 56127 Pisa, Italy}

\begin{abstract}
There is a great interest in the scientific community to perform calorimetry on samples having mass in the nanogram range. A detailed knowledge of the energy (heat) exchange in the fast growing family of micro-- and nano--systems could provide valuable information about the chemistry and physics at the nano--scale. The possibility to have an atomically flat thermal probe represents an added value, because it provides the unique opportunity to perform Scanning Probe Microscopy (SPM) together with calorimetry. Here we report the fabrication, characterization, and calibration of atomically flat, single--crystalline gold film thermometers on mica substrate. Gold re--crystallization has been obtained, and successively the thermometer surface has been studied by Low Energy Electron Diffraction (LEED) and Scanning Tunneling Microscopy (STM). The thermometer calibration demonstrates a heat exchange coefficient of  $2.1 \times 10^{-7}$~W/K and a performance about 10 times better than previous sensors based on Si substrates. The experimental setup allows the simultaneous investigation of heat exchange and surface physics  on the same sample.

\end{abstract}




\end{frontmatter}



\section{Introduction}

Calorimetry is a powerful tool to investigate the heat exchange between a sample and its environment. The technique is widely used and has allowed a systematic study of phase transitions, chemical reactions, and in general all processes which involve the modification of a sample's structure or its thermodynamic conditions.  A detailed knowledge of the energy balance during the evolution of a system provides an invaluable insight on the properties of a sample and on the driving forces acting on it. In the last decade, a growing number of devices and sensors at the micro-- and nano--scale were developed, pushing calorimetry to follow the trend. However, commercial calorimeters, up to now, can only deal with samples having masses in the mg range and have an energy sensitivity of about 1 mJ. Even if performance is improving with time, commercial calorimeters are still far from access to nano--scale samples. A calorimetric evaluation of tiny samples would represent a valuable source of information on molecules, for example, of pharmaceutical interest, as decreasing sample mass reduces considerably the development expenses of drugs.

Recently, we have demonstrated an original calorimetric technique based on a gold film thermometer~\cite{Basta18}. The experimental set--up allowed the measurement of the heat released during the adsorption of a sub--nanogram amount of hydrogen on Ti--functionalized monolayer graphene. The total sample mass was about 10 ng, and the amount of adsorbed hydrogen a fraction of a nanogram. The amount of material necessary to perform the measurement was about $5-6$ orders of magnitude lower than what commercial calorimeters request. The gold film was deposited on an amorphous silica (SiO$_2$) layer, on top of a silicon substrate, which resulted in a thermometer surface which is rough at the atomic level~\cite{Basta18}. This surface roughness prevents the possibility to perform STM analysis with atomic resolution. To achieve this result, which would allow to obtain surface information from the sample under investigation, an atomically flat thermal probe is required.

Thin metal films offer several advantages over metal single crystals for scanning probe microscopy experiments. Thin films can be directly prepared in vacuum, thereby avoiding the possibility of contamination from polishing or transfer through atmosphere, and the cost is usually a small fraction of the corresponding single crystal~\cite{Dishner98}. For many research and industrial applications, the preparation of high quality gold thin films is essential~\cite{Buchholz91,Golan92,Porath94}.

For this purpose, mica is often used as substrate. Its two--dimensional structure, transparency, elasticity, flexibility, and chemical inertness make mica suitable for applications. Mica is additionally inexpensive and abundant, lightweight and biocompatible~\cite{McCoul16}. All these features qualify mica as an optimal substrate for flexible electronics~\cite{Wong09}. The mica properties most relevant for this investigation are its low thermal conductivity, which shows an anisotropic behaviour (4.05~W/m$\cdot$K parallel to the cleavage planes, and 0.46~W/m$\cdot$K perpendicular to the cleavage planes~\cite{Gray77}, to be compared with Si, 156~W/m$\cdot$K)~\cite{Glassbrenner64}, and its capability to allow gold epitaxial re--crystallization along (111) crystallographic planes~\cite{Dishner98}.

Upon thermal annealing, the Au(111) surface arranges in the so--called herringbone reconstruction~\cite{Barth90}. The herringbone reconstruction consists of pair--wise arranged parallel lines running in a zigzag pattern. Atomic resolution images of the surface have allowed the direct determination of the unit cell of the reconstructed layer~\cite{Barth90}. The lattice vectors are given by the connection between adjacent main minima (63~\AA{} in the $[1\bar{1}0]$ direction), and by the connection between next--neighbor $[1\bar{1}0]$ rows of Au atoms (4.7~\AA{} in the $[11\bar{2}]$ direction), thus forming a unit cell of 63~\AA{} $\times$ 4.7~\AA, called $22\times \sqrt{3}$ herringbone reconstruction~\cite{Harten85,Woll89,Barth90}.

A gold thin film thermometer on a mica substrate therefore offers the opportunity to fabricate an atomically flat thermometer, which besides calorimetry applications allows to utilize scanning probe microscopy in order to carry out more complex surface physics experiments. Another positive side effect of the use of mica as a substrate is represented by the fact that mica, depending on the direction, has a  40 to 300 times lower thermal conductivity than silicon~\cite{Glassbrenner64,Gray77}.

Here, we report on the development of a sensitive, atomically flat gold film thermometer on mica substrate. We first describe the thermometer's surface characterization showing a reconstructed gold surface. We demonstrate resolution at the \AA ngstrom scale during STM imaging of the surface of the thermometer.  Moreover, the performance of the Au/Mica thermometers reported in the present work is about one order of magnitude better than that obtained with Si substrates.  Our work opens the possibility to simultaneously investigate  energy (heat) exchange mechanisms and surface physics with the same physical support. Indeed, the possibility to obtain calorimetric information as well as surface characterization with atomic resolution on the same sample in a clean and controlled UHV environment opens a unique perspective in understanding physics and chemistry at the nano--scale.

\section{Experimental}

In the present investigation, two series of samples have been produced, in order to demonstrate the reproducibility of the fabrication process. Both series have been realized evaporating 20~nm of gold directly on an approximately 250~$\mu$m--thick, freshly cleaved mica substrate. Before gold deposition, the mica has been annealed at $200^\circ$C overnight under high vacuum ($3 \times 10^{-6}$ mbar). Then, the mica was cut into $\sim 5 \times 5$~mm$^2$ samples with scissors. More details on sample fabrication and structure are reported in the Supporting Information.

All experiments on thermometer characterization and calibration have been performed in an ultra--high vacuum (UHV) environment, inside a RHK Technology VT--STM chamber. Sample heating can be performed in two different ways: utilizing a silicon slab under the mica as heater (degassing and thermometer calibration) or illuminating the sensor with a white light source able to deliver a power of 13 mW to the sample (heat exchange coefficient measurements). Each sample has been degassed overnight at a temperature between $100^\circ$C and  $140^\circ$C and successively annealed at $200^\circ$C in UHV. The heating ramps were controlled via PC using a dedicated LabVIEW software. The resistance of the gold film thermometers is measured with a Wheatstone bridge cascaded to a high quality preamplifier, the output of which is recorded by a lock--in amplifier (Stanford Research Systems SR830).

For a complete analysis, we performed LEED measurements and took STM images of different areas of the gold layer of each sample. STM imaging was obtained at room temperature with a base pressure around  $2 \times 10^{-11}$ mbar. Further details on the experimental setup are reported in the Supporting Information.

\section{Results and discussion}

\subsection{\label{sec:morph}Morphological characterization of the Au film thermometers}

\begin{figure}[tbh]
\centering
   \includegraphics[width=0.7\columnwidth]{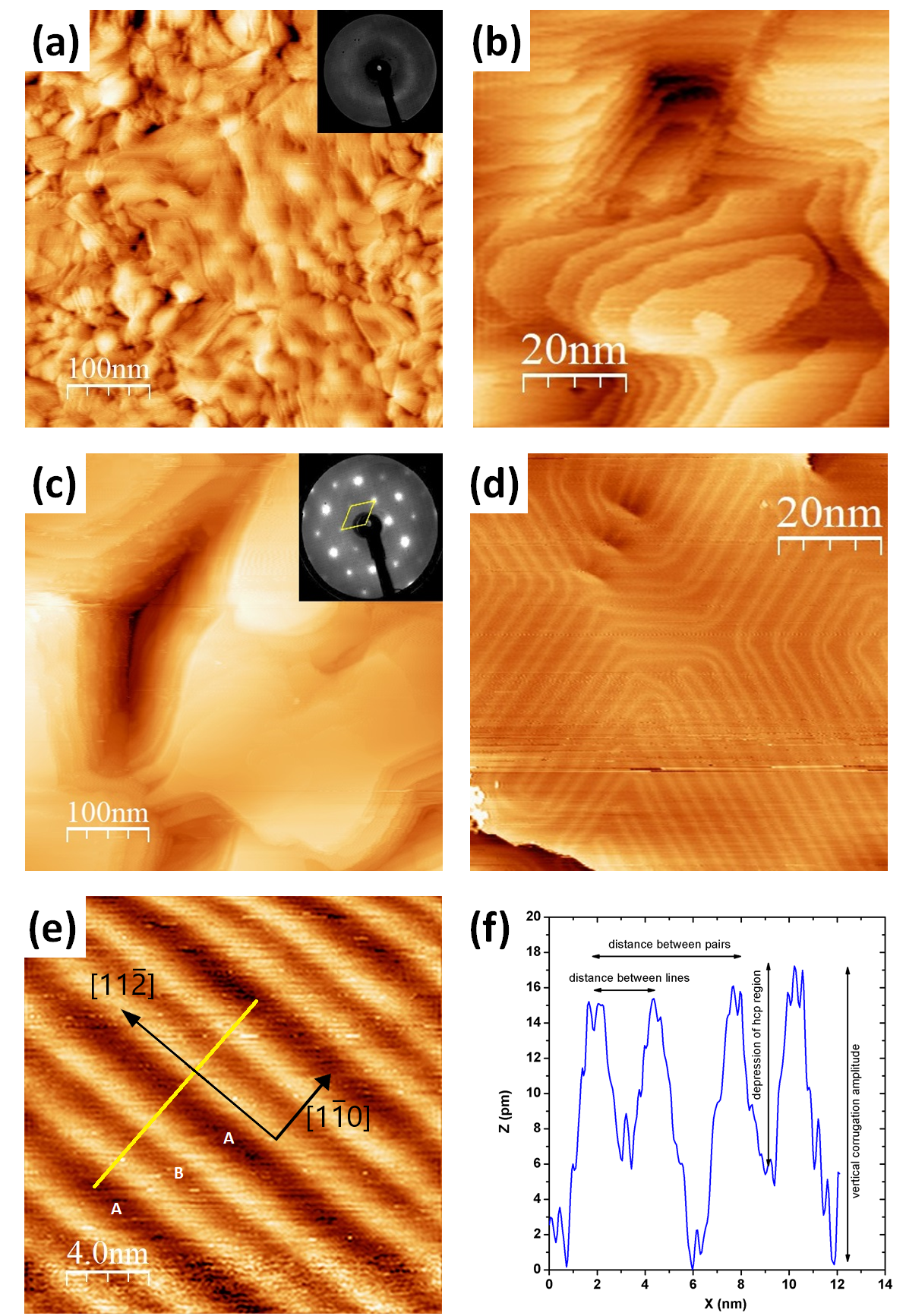}
   \caption{\label{M2-5_100} (a,b) STM images from a gold thermometer sample after a heating ramp  to $100^\circ$C. Image size: (a) $500 \times 500$~nm$^2$, LEED pattern in the inset  (electron energy 103.5 eV), (b) $80 \times 80$~nm$^2$. (c-e) STM images from a gold thermometer sample after a heating ramp  to $200^\circ$C. Image size: (c) $500 \times 500$~nm$^2$, LEED pattern in the inset (electron energy 103.5 eV), the yellow diamond represents the unit cell, (d) $80 \times 80$~nm$^2$, and (e) $20 \times 20$~nm$^2$. (d) shows the corrugation lines typical for the gold surface herringbone reconstruction. Letters in (e) label the fcc (A) and hcp (B) regions. Crystallographic directions are indicated. (f) Section view of the corrugation lines, taken along the yellow line in (e).}
\end{figure}

After degassing the gold--on--mica samples, LEED shows a broad, diffuse diffraction pattern (inset to Fig.~\ref{M2-5_100}(a)), indicating a random orientation of the gold grains. STM images, shown in Figs.~\ref{M2-5_100}(a) and (b), are in good agreement with the LEED results. The gold surface consists of rolling hills with few atomically flat regions and narrow terraces, about $2-10$~nm wide. Stable and reproducible images were obtained by setting 1.3~V as bias voltage and $0.7-0.8$~nA as set point current. From the LEED and STM results, it is clear that a temperature of $100^\circ$C is not sufficient to achieve re--crystallization. The sample has therefore been annealed up to $200^\circ$C, with a rate of $1^\circ$C/minute following the procedure reported in Ref.~\cite{Nan97}. After this, the LEED analysis reveals bright spots arranged in a hexagonal pattern which characterizes the (111) gold surface (see inset to Fig.~\ref{M2-5_100}(c)). This result is in good agreement with literature, which reports that gold re--crystallization is achieved at temperatures ranging from $170^\circ$C to $200^\circ$C~\cite{Nan97}. LEED analysis has been performed in several positions, mapping the whole surface. The LEED data unequivocally demonstrate that the gold film is single-crystalline.

Figure~\ref{M2-5_100}(c) shows the topography of the clean and well--annealed Au(111) surface after this heating ramp. It is characterized by large terraces, which have a width of up to 300~nm and are atomically flat, in agreement with literature~\cite{Harten85,Woll89,Barth90,Nan97}, with an average rms roughness of 6.2 pm over a surface area of 400~nm$^2$. This allows high resolution STM imaging and opens the possibility to perform surface physics experiments on the thermometer itself. 

This can be appreciated in Figs.~\ref{M2-5_100}(d) and (e), which show that the terraces exhibit the well--known periodic pattern of pairwise--arranged, parallel lines running in $[11\bar{2}]$ directions which characterize the herringbone reconstruction and separate fcc from hcp regions~\cite{Barth90}. Moreover, U--shaped connections between neighboring corrugation lines are clearly visible in Fig.~\ref{M2-5_100}(d). The three different rotational domains, which characterize the herringbone reconstruction, are observed. The corrugation line pairs are characteristically deformed in the vicinity of the bending points, in agreement with literature~\cite{Barth90}. We observe not only the elbows typical of the herringbone pattern, but also linear or even more complex patterns. This behavior has been explained in terms of compressive strain by the mica substrate~\cite{Narashimhan92,Schaff01}.

Figure~\ref{M2-5_100}(f) shows a cross--section along the $[1\bar{1}0]$ direction taken along the yellow line  in Fig.~\ref{M2-5_100}(e). The average distance between neighboring pairs in $[1\bar{1}0]$ direction amounts to $(5.6 \pm 0.7)$~nm, while the individual lines within a pair are $(2.4 \pm 0.2)$~nm apart. The average corrugation amplitude amounts to $(18 \pm 5)$~pm, while the depression of the narrow regions between the two lines of each pair is less pronounced with $(12 \pm 5)$~pm, in agreement with literature values~\cite{Barth90, Woll89}.

\subsection{\label{sec:sim}Simulation of thermometer performance}

In order to quantify the performance of the device (thermometer plus readout) to detect a thermal signal, we have simulated the heating of the thermometer with a uniform energy release at the top of the device. In the model, we consider the sensor structure and the related characteristic times. The main contributions are connected to the thermalization time of the gold film, the thermal resistance at the interface between the gold film and mica, the thermalization time of the mica substrate, and the thermal resistance at the interface between mica and sample holder. While the thermalization times of gold film and mica substrate are easy to evaluate, the heat transfer rates at the interfaces (Au/mica and mica/sample holder) are more difficult to assess.

Heat diffusion inside the gold film is very fast. The time scale for heat redistribution in a 20 nm gold layer is $t_{Au} \sim d^2/\alpha \sim 4 \times 10^{-12}$~s, with $d$ the thickness of the gold film and $\alpha$ the gold thermal diffusivity~\cite{Wolf04,Carl59}. Therefore we can assume that the gold film is at uniform temperature and leaks heat towards the mica substrate. Similarly, the time scale for heat redistribution in a 250~$\mu$m mica substrate is $\sim 0.33$~s~\cite{Gray77}. Overall, the thermalization between gold and mica will have a characteristic time $\tau$ greater than 0.33~s due to the additional contribution of the thermal resistance at their interface, but since this interface contribution is difficult to assess, we assume a characteristic time $\tau = 0.33$~s, which is a lower bound of the real value of $\tau$.

We note that this characteristic time fits well the experiments for which we have optimized the thermometer. A chemical reaction occurring in UHV environment between a functionalized surface and molecules in the gas phase usually runs slowly. At a gas supply pressure $p$ lower than $10^{-7}$ mbar, the time necessary to have a monolayer of molecules  on the surface ranges from several seconds to minutes (about 10 s at $p = 10^{-7}$~mbar, about 100 s at $p = 10^{-8}$~mbar)~\cite{Oura13}.

\begin{figure}[tbh]
\centering
   \includegraphics[width=0.7\columnwidth]{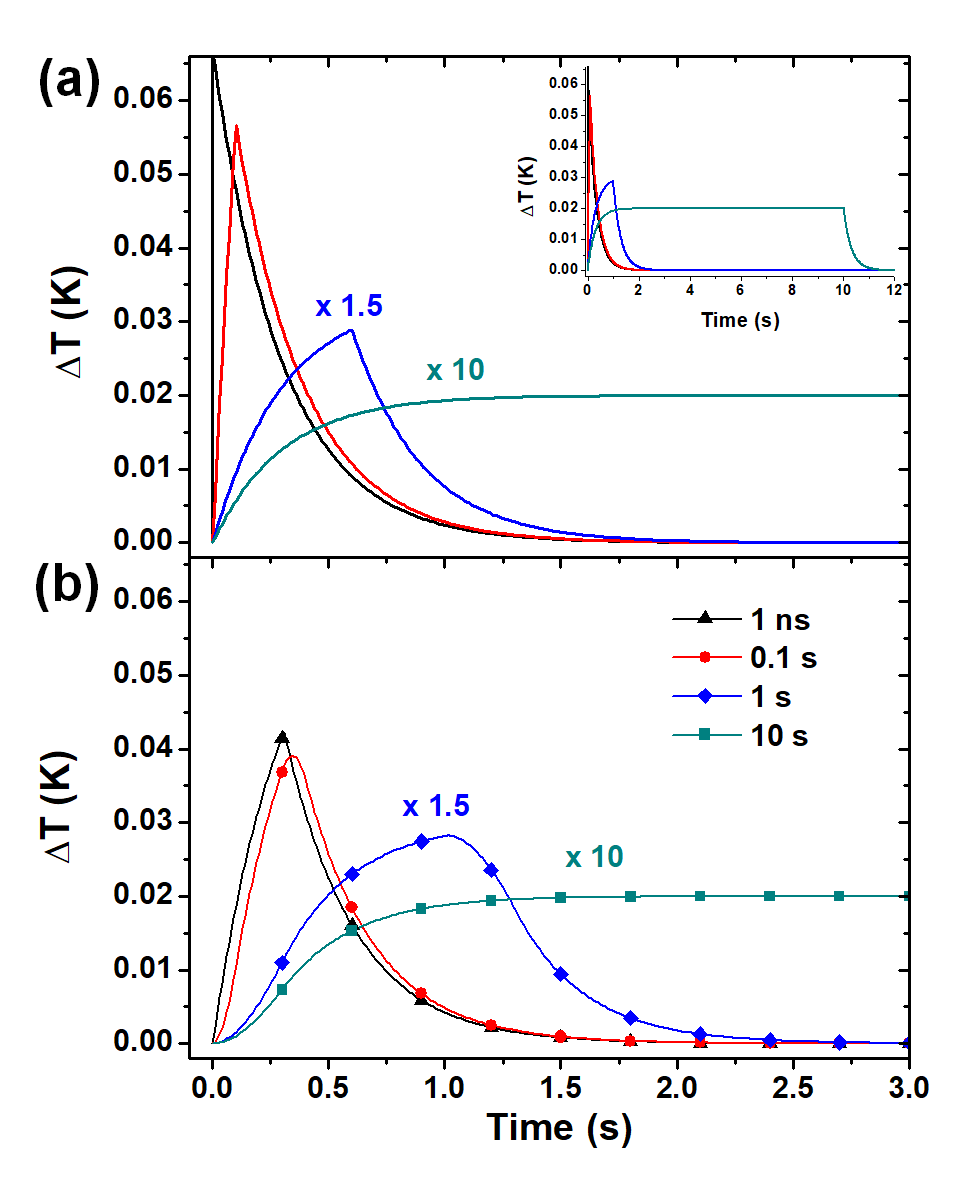}
   \caption{\label{Sim} (a) Variation in thermometer temperature vs.~time after a rectangular heating pulse of width 1 ns, 0.1 s, 1 s, and 10 s. Lines represent the calculated temperature, inset shows a wider view (up to 12 s). (b) Temperature variation (symbols) as it would be measured by a data acquisition system with integration time $\tau_i = 300$~ms.}
\end{figure}

Thus, we have calculated the variation in thermometer temperature due to rectangular heating pulses of various durations, ranging from much shorter ($10^{-9}$~s) to much longer (10~s) than the characteristic time $\tau= 0.33$~s. Details are provided in the Supporting Information. The results of the simulation are reported in Fig.~\ref{Sim}(a), which shows the variation in thermometer temperature with time when rectangular heating pulses of duration 1~ns, 0.1~s, 1~s, and 10~s are applied.

In experiment, the resistance of the gold film thermometer is measured with a lock--in amplifier with an integration time $\tau_i$ (typically $\tau_i = 300$~ms). This introduces another time scale in the problem which we have simulated with a boxcar averaging of the temperature data, with the boxcar width corresponding to the lock--in integration time. Fig.~\ref{Sim}(b) shows the temperature signal (cf.~Fig.~\ref{Sim}(a)) as it would be measured by the readout electronics. The line is obtained applying a boxcar to the simulated temperature data, while the symbols indicate data points measured every 300~ms.

The analysis of such temperature profiles, measured or simulated, has been described in detail in Ref.~\cite{Basta18}. In brief, the system is described by a simple one--dimensional thermal model, in which the thermometer is heated by the absorption of a thermal power $P(t)$ while at the same time it releases energy via heat losses towards the substrate. The losses are described by the heat exchange coefficient $\lambda$. Assuming that the heat capacity of the sensor remains constant during the measurement, we can write\cite{Barbini89,Bertolini90,Cassettari93,Basta18}
\begin{equation}
   \frac{\delta H_r}{\delta t}=C_{sensor}\cdot\frac{\delta\Delta T(t)}{\delta t}+\lambda\cdot\Delta T(t), \label{5}
\end{equation}
with $H_r$ the total heat released, $\Delta T(t)$ the resulting temperature increase of the sensor, and $C_{sensor}$ the heat capacity of the sensor at constant pressure. We consider the temperature signal from the start of heating to the maximum of the temperature curve. Applying this procedure to the simulated signals shown in Fig.~\ref{Sim}(b) allows to calculate the energy which has been supplied to the system and which we label $E_c$. Here, this energy is perfectly known because it is an input parameter of the simulations. We label it $E_s$. A comparison between $E_c$ and $E_s$ allows to judge the quality of the algorithm used for data analysis and to estimate the related error. For this purpose, we calculate the ratio between the difference of the supplied and calculated energy $\Delta E = E_s - E_c$ and the supplied energy $E_s$. Results are summarized in Table~\ref{simu}.

\begin{table}[tbh]
\centering
 	\caption{\label{simu} Heating pulse duration and relative error $\Delta E/E = (E_s - E_c)/E_s$ in energy, where $E_s$ is the supplied energy and $E_c$ the calculated one following the procedure reported in Ref.~\cite{Basta18}.}
  \begin{tabular*}{0.8\columnwidth}{@{\extracolsep{\fill}}ccc} 
  	\hline
    \textbf{Heating Pulse} & \textbf{Integration} & \textbf{$\Delta$E/E}  \\
    \textbf{Duration [s]} & \textbf{Time $\tau_i$ [ms]} & \\
	  \hline
    $10^{-9}$ & 300 & 0.04  \\
    0.1 & 300 & 0.053   \\
    1 & 300 & 0.007  \\
    10 & 300 & 0.005  \\ 
    \hline
  \end{tabular*} 
\end{table}

The simulations show a good agreement between the applied energy (the integral under the heating pulse) and the calculated energy using Eq.~\ref{5}, and therefore validate the model. At first sight it might seem surprising that good agreement is obtained even for heating pulses much shorter than the characteristic times of the system. However, this can be understood as follows: the fast heating pulse leads to an almost immediate temperature increase in the gold film, on a time scale much faster than the integration time of the electronics. This energy is stored in the gold film and can be dispersed only within the characteristic time of the mica substrate ($\sim 1$~s), which is well within the accessible range of the electronics. In other words, while the dynamics of fast heating pulses cannot be resolved by the present electronics, the integral amount of the deposited energy can be measured with a high precision. We note that even fast processes could be followed dynamically with a fast electronics: commercially available lock--in amplifiers offer a lowest time constant of 30~ns for demodulation, seven orders of magnitude faster than the electronics employed for our experiments.

\subsection{Thermometer calibration}

\begin{figure}[tbh]
\centering
   \includegraphics[width=0.7\columnwidth]{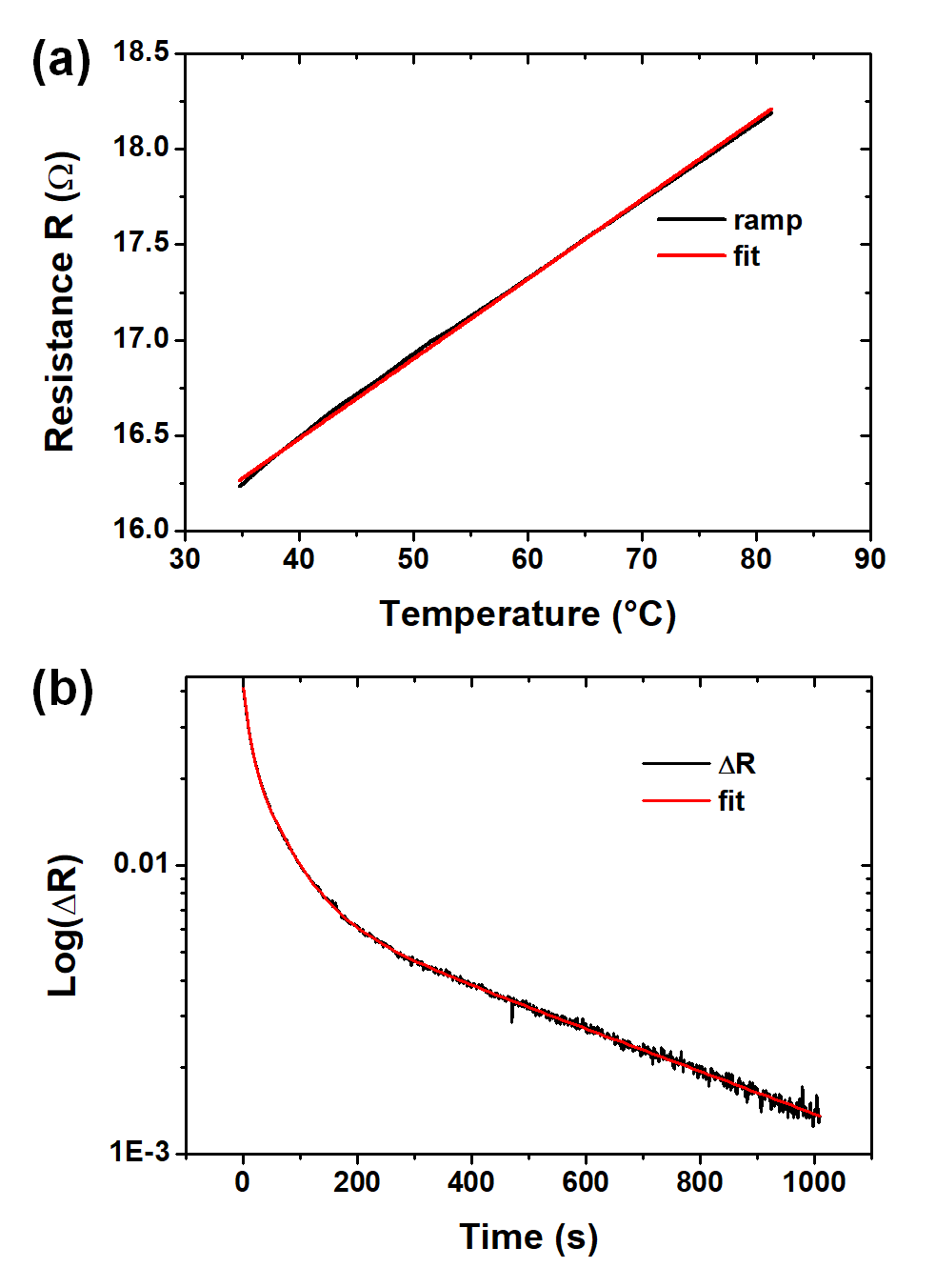}
   \caption{\label{M5-4_ramps} (a) Data and linear fit of a calibration ramp relative to sample M6-1. (b) Sample thermalization after an applied heating pulse. Log--plot of experimental data in black, while the exponential fit is shown in red.}
\end{figure}

A thermometer calibration protocol has been applied to two samples. We carried out the calibration measurements applying slow ramps from room temperature  to  $100^\circ$C. The heating and cooling rates of the calibration ramps were about $2.8^\circ$C/s. During the heating ramps we recorded the temperature of the gold sensor as measured by a thermocouple, the current and the voltage supplied to the heater, and the sample resistance. A plot of resistance $R$ vs.~temperature $T$ is shown in Fig.~\ref{M5-4_ramps}(a). From the calibrations, the temperature coefficient and the electrical resistivity of the gold thermometers are obtained and can be compared with literature.

As expected, there is a linear proportionality, and therefore we have performed a linear fit of the data using the relation
\begin{equation}
R(T)=R_0[1+\alpha(T-T_0)]. \label{law}
\end{equation}
The intercepts obtained from the fits correspond to $R_0 \equiv R(T_0)$, while the slopes correspond to the product $R_0\cdot\alpha$. The reference temperature $T_0$ is the temperature measured by the thermocouple just before the start of the calibration ramp. The parameters obtained for the two samples M5-4 and M6-1 are listed in Table~\ref{M5-4_ramps_tab}. The average value is the result of a weighted average, and the given error is the standard deviation of the average.

\begin{table}[tbh]
\centering
 	\caption{\label{M5-4_ramps_tab}Parameters of the linear fits and the weighted averages. T.C.R. stands for Temperature Coefficient of Resistance.}
  \begin{tabular*}{0.95\columnwidth}{@{\extracolsep{\fill}}cccccc}
  	\hline
    \textbf{Sample} & \textbf{Ramp} & \textbf{T.C.R.} & \textbf{Error} & \textbf{Intercept} & \textbf{Error} \\
    & & \textbf{($\alpha$) [K$^{-1}$]} & & \textbf{($R_0$) [$\Omega$]} & \\
	  \hline
    M5-4 & 1 & 3.06$\cdot10^{-3}$ & 5$\cdot$10$^{-5}$ & 12.208 & 0.001  \\
    M5-4 & 2 & 3.18$\cdot10^{-3}$ & 8$\cdot$10$^{-5}$ & 12.092 & 0.003  \\
    M6-1 & 1 & 2.78$\cdot10^{-3}$ & 5$\cdot$10$^{-5}$ & 14.910 & 0.001  \\
    M6-1 & 2 & 2.80$\cdot10^{-3}$ & 5$\cdot$10$^{-5}$ & 14.832 & 0.001  \\
    M6-1 & 3 & 2.82$\cdot10^{-3}$ & 5$\cdot$10$^{-5}$ & 14.813 & 0.001  \\
    M6-1 & 4 & 2.86$\cdot10^{-3}$ & 6$\cdot$10$^{-5}$ & 14.786 & 0.002  \\
    M6-1 & 5 & 2.87$\cdot10^{-3}$ & 6$\cdot$10$^{-5}$ & 14.786 & 0.002  \\ 
    \hline
	  Average && 2.91$\cdot10^{-3}$ & 1.5$\cdot$10$^{-4}$ & &  \\
	  \hline
  \end{tabular*}
\end{table}

The values of $\alpha$ for M5-4 and M6-1 are consistent within a standard deviation. The overall reproducibility is very good, within 5\%.  A value of $\alpha$ lower than the tabulated literature bulk value ($\alpha = 3.4 \times 10^{-3}$~K$^{-1}$)\cite{Serway12} is expected because the thickness of the gold film (20~nm) is lower than its electron mean free path $\lambda \sim 40$~nm~\cite{Lacy2011,Gall16}. However, the thin film contributions to thermometer resistance do not affect the validity of the procedure, because we perform a complete calibration of each device before its use as a sensor. From the noise level in the resistance measurements, $\delta R \sim 1.54 \times 10^{-4}$~$\Omega$, we can determine the sensitivity of the thermometer, $T_n = \delta R / \left( \alpha R_0 \right) \sim 4$~mK, which is about 2 times better than the previously reported sensor based on a Si substrate~\cite{Basta18}.

In order to analyze calorimetric data using Eq.~\ref{5}, we need to determine the sensor heat capacity $C_{sensor}$ and the heat exchange coefficient $\lambda$. In particular, for $\lambda$ we can expect three different contributions: since the mica substrate acts as a thermal insulator, we expect a first and fast thermalization of the upper gold film with the underlying mica substrate, followed by a slower thermalization of the substrate with the sample holder, and finally a very slow thermalization of the sample holder with the environment (see Supporting Information for details). In order to disentangle the three contributions, we have applied a heating step ($\sim$ 60 s duration) to the sample and measured the temperature decay, as shown in Fig.~\ref{M5-4_ramps}(b) (see Supporting Information for details). A fit of this curve with the sum of three exponential decay functions yields the characteristic thermalization times. In particular, the fast  thermalization time between gold and mica is  $\tau_1=(10.8\pm0.1)$~s.

To determine the heat exchange coefficient $\lambda$ from the decay time $\tau_1$, we need to evaluate the sample heat capacity. Since the gold film thermalizes very fast with respect to the mica, we can assume that the sensor heat capacity is due to the gold film only. From the gold film thickness and the area of the sensor, its heat capacity result to be
\begin{equation}
C_{sensor} = \rho_{Au} \cdot t_{Au} \cdot A_{Au} \cdot c_{Au} = 2.24 \times 10^{-6} \, \mathrm{J/K,}
\end{equation}
with gold density $\rho_{Au}=19.3$ g/cm$^3$, sample thickness $t_{Au} = 20$~nm, sample dimensions 6.90~mm $\times$ 6.75~mm, and gold specific heat capacity $c_{Au}=0.13$ J/K$\cdot$g. Next, we can calculate the sensor heat exchange coefficient
\begin{equation}
\lambda = C_{sensor} / \tau_1 = (2.07 \pm 0.02) \times 10^{-7} \, \mathrm{W/K.}
\end{equation}
If we compare this result with the $\lambda$ measured for the sensors based on Si substrate~\cite{Basta18}, we have obtained an improvement by a factor 25. The equivalent noise power $P_n$ of the sensor is defined as the product of the thermometer heat exchange coefficient $\lambda$ and the thermometer sensitivity $T_n$,
\begin{equation}
P_n = \lambda \cdot T_n,
\end{equation}
which gives a value of 1.0 nW for the gold--on--mica sensor, to be compared to the value of 50 nW for the gold--on--Si sensor, i.e.~an improvement by a factor of 50.

\begin{figure}[tbh]
\centering
   \includegraphics[width=0.7\columnwidth]{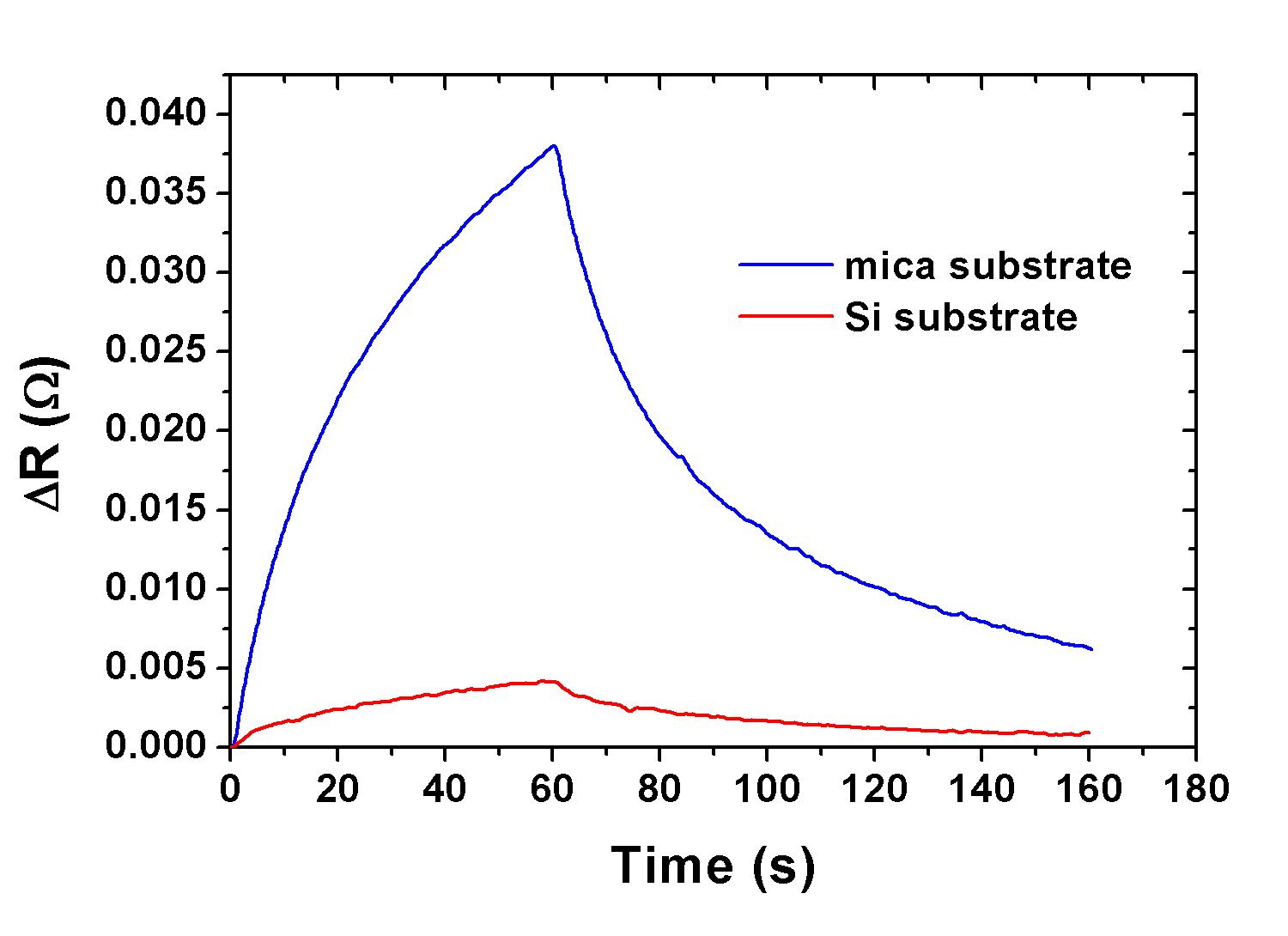}
   \caption{\label{ExpDec3}Response of the two sensors (gold--on--mica and gold--on--Si) to illumination with a lamp for 60 seconds.}
\end{figure}

To further compare the performance of the two sensors, based on Si and on mica substrate, each sensor has been subjected to identical heating pulses. In particular, the sensors were exposed to an illumination with a white light lamp for 60~s. The results are shown in Fig.~\ref{ExpDec3}. It is clearly evident that the resistance increase measured with the gold--on--mica sensor ($\Delta R_{mica}=0.038$~$\Omega$) is almost an order of magnitude larger than that of the gold--on--Si sensor ($\Delta R_{Si}=0.004$~$\Omega$). This result experimentally confirms that the sensitivity achieved with the new set--up is significantly higher compared to the previous sensors based on Si. The improvement pushes the sensitivity in the exchanged heat detection to $\Delta E$ better than 0.5~$\mu$J. Moreover, we can consider this evaluation an underestimation, because the gold film thickness is small with respect to the light wavelength in the visible region. Therefore light is partially transmitted through the gold film. The transmitted fraction is completely absorbed by Si, while it is almost completely transmitted by the mica which is transparent for visible light.

 \bibliographystyle{elsarticle-num} 
  \bibliography{bib}





\end{document}